\documentclass{PoS}

\usepackage{amsmath}
\usepackage{booktabs}
\usepackage{subcaption}

\DeclareMathOperator{\sign}{sign}

\title{Numerical study of ADE-type $\mathcal{N}=2$ Landau--Ginzburg models}

\ShortTitle{Numerical study of ADE-type $\mathcal{N}=2$ Landau--Ginzburg models}

\author{\speaker{Okuto Morikawa}
        \\
        Department of Physics, Kyushu University, 744 Motooka, Nishi-ku, Fukuoka, 819-0395, Japan\\
        E-mail: \email{o-morikawa@phys.kyushu-u.ac.jp}}

\abstract{%
At an extremely low-energy scale,
it is believed that the two-dimensional~$\mathcal{N}=2$ Wess--Zumino model becomes
an~$\mathcal{N}=2$ superconformal field theory (SCFT).
We study this theoretical conjecture of the Landau--Ginzburg (LG) description
by numerical simulations based
on a supersymmetric-invariant momentum-cutoff regularization.
First, from the two-point function of the energy-momentum tensor,
we measure the central charge of the ADE minimal models.
Second, we develop a method to take the continuum limit,
and perform a precision measurement of the scaling dimension
in the~$A$-type minimal model.
All our results show a coherence picture being consistent
with the conjectured LG/SCFT correspondence.
}

\FullConference{37th International Symposium on Lattice Field Theory - Lattice2019\\
		16-22 June 2019\\
		Wuhan, China}

\begin{document}

\section{Introduction}
It is believed that
the two-dimensional ($2$D) massless $\mathcal{N}=2$ Wess--Zumino (WZ) model
with a quasi-homogeneous superpotential
provides a Lagrangian-level realization
of the~$2$D~$\mathcal{N}=2$ superconformal field theory (SCFT).
Such a SCFT would be a scale-invariant theory
on the nontrivial infrared (IR) fixed point of the WZ model,
while all massive modes are decoupled.
This conjecture of the Landau--Ginzburg (LG) description
has been theoretically analyzed from various aspects;
e.g., see~\cite{Witten:1993jg}.
To a solvable ADE-type minimal model,
the corresponding superpotential of the WZ model is shown
in~table~\ref{tab:ade}~\cite{Vafa:1988uu}.
It is, however, difficult to prove this theoretical conjecture directly,
since the coupling constant becomes strong at the IR region
and the perturbation theory possesses IR divergences.
The LG description is remarkably a non-perturbative phenomenon.

\begin{table}[h]
 \centering
 \begin{tabular}{lll}\toprule
  Algebra & Superpotential $W$                         & Central charge $c$ \\\midrule
  $A_n$   & $\Phi^{n+1}$, $n\geqq1$                    & $3-6/(n+1)$        \\
  $D_n$   & $\Phi^{n-1}+\Phi\Phi^{\prime2}$, $n\geqq3$ & $3-6/2(n-1)$       \\
  $E_6$   & $\Phi^3+\Phi^{\prime4}$                    & $3-6/12$           \\
  $E_7$   & $\Phi^3+\Phi\Phi^{\prime3}$                & $3-6/18$           \\
  $E_8$   & $\Phi^3+\Phi^{\prime5}$                    & $3-6/30$           \\
  \bottomrule
 \end{tabular}
 \caption{ADE classification~\cite{Vafa:1988uu}}
 \label{tab:ade}
\end{table}

An alternative approach to this issue may be provided
by a non-perturbative calculational method such as the lattice field theory.
This kind of numerical method, when further developed,
may enable us to compute directly scattering amplitudes
in a superstring theory whose world sheet theory is given by an~$\mathcal{N}=2$ SCFT;
the theory possesses the superstring compactification
to the Calabi--Yau quintic threefold.
Such a theory is in general not a minimal model nor a product of minimal models.
With regard to this point,
the LG description realizes a specific strongly-interacting Lagrangian corresponding
to the Calabi--Yau manifold~\cite{Greene:1988ut,Witten:1993yc}.
A numerical approach to the~$2$D~$\mathcal{N}=2$ WZ model
would be useful to investigate a superstring theory.

As is well recognized, however,
the lattice regularization is generally incompatible with the supersymmetry (SUSY).
The lattice parameters should be fine-tuned
so that a lattice model yields the target SUSY continuum theory.
To this issue, a possible solution is that
we construct the~$2$D~$\mathcal{N}=2$ WZ model on the lattice
on the basis of the so-called Nicolai map~\cite{Nicolai:1979nr,Nicolai:1980jc}.
For example, in the lattice formulation from~\cite{Kikukawa:2002as},
one nilpotent SUSY is exactly preserved at finite lattice spacing,
and the vacuum energy is canceled even on the lattice owing to the lattice Nicolai map.
Moreover, it can be argued that, to all orders of perturbation theory,
the full SUSY is automatically restored in the continuum limit without any fine tuning.
By using this formulation~\cite{Kikukawa:2002as},
the scaling dimension of the scalar field in the~$A_2$-type theory
with the cubic superpotential was measured~\cite{Kawai:2010yj}.
This numerical study achieved a triumph of the lattice field theory.

Somewhat later,
the authors in~\cite{Kamata:2011fr} examined the same~$A_2$-type WZ model
by using the formulation from~\cite{Kadoh:2009sp},
and measured the scaling dimension and the central charge.
The formulation~\cite{Kadoh:2009sp} is based on the Nicolai mapping
and the momentum cutoff regularization,
and preserves the full set of SUSY
as well as the translational invariance even with a finite cutoff.
Then, the construction of the Noether currents associated with spacetime symmetries,
e.g., the supercurrent and the energy-momentum tensor (EMT), is straightforward.
This feature enables us to compute the central charge,
which appears in two-point functions of such Noether currents.

In this paper,
we numerically study the~$2$D~$\mathcal{N}=2$ WZ model,
based on the momentum-cutoff regularization~\cite{Kadoh:2009sp}.
First, we focus
on the $A_2$, $A_3$, $D_3$, $D_4$, $E_6$ ($\cong A_2\otimes A_3$), and $E_7$ models.
The method in~\cite{Kamata:2011fr} is generalized
to the WZ model with multiple superfields and more complicated superpotentials.
From the IR behavior of the EMT correlator,
we numerically determine the central charge of
these models~\cite{Morikawa:2018ops,Morikawa:2018zys}.
Second, we develop an extrapolation method
to take the continuum and infinite-volume limit~\cite{Morikawa:2019cql},
while any extrapolation has been not done in the preceding numerical studies.
Then, on the basis of the formulation~\cite{Kadoh:2009sp},
we perform a precision measurement of the scaling dimension in the~$A_2$-type theory.
Our results below show a coherence picture being consistent
with the conjectured LG description of the ADE minimal models.

\section{SUSY-preserving formulation using the Nicolai map}
First of all, we briefly review the SUSY-preserving formulation in~\cite{Kadoh:2009sp}.
In what follows,
the system is defined in a~$2$D Euclidean physical box~$L_0\times L_1$;
let us work in the momentum space with a momentum cutoff, $p_\mu = 2\pi n_\mu/L_\mu$
($n_\mu=0$, $\pm1$, \dots, $\pm L_\mu/2a$),
where the Greek index~$\mu$ runs over~$0$ and~$1$,
and repeated indices are not summed over.
Here, $a$ is a unit of dimensionful quantities;
the \emph{continuum limit} $a\to0$ removes the UV cutoff.
For simplicity, we take $L/a=L_0/a=L_1/a$ as even integers.

Let us consider the $2$D~$\mathcal{N}=2$ WZ model
with~$N_{\Phi}$ supermultiplets, $\{\Phi_I\}_{I=1,\dots,N_{\Phi}}$,
which consist of complex scalar fields~$\{A_I\}$,
and left- and right-handed spinors~$\{(\psi_\alpha, \Bar\psi_{\Dot{\alpha}})_I\}$
($\alpha=1$, $2$).
Then, the action of the~$2$D~$\mathcal{N}=2$ WZ model
with a quasi-homogeneous superpotential~$W(\{A\})$ is given by
\begin{align}
 S &= \frac{1}{L_0L_1} \sum_p \sum_I\Biggl[
 4 p_z A_I^*(-p) p_{\Bar{z}} A_I(p)
 + \frac{\partial W(\{A\})}{\partial A_I}(-p)
 \frac{\partial W(\{A\})^*}{\partial A_I^*}(p)
 \notag\\&\qquad\qquad\qquad\qquad
 + (\Bar\psi_{\Dot{1}}, \psi_2)_I(-p) \sum_J
 \begin{pmatrix}
  2\delta_{IJ}p_z & \frac{\partial^2 W(\{A\})^*}{\partial A_I^* \partial A_J^*}* \\
  \frac{\partial^2 W(\{A\})}{\partial A_I \partial A_J}* & 2\delta_{IJ}p_{\Bar{z}}
 \end{pmatrix}
 \begin{pmatrix}
  \psi_1 \\ \Bar\psi_{\Dot{2}}
 \end{pmatrix}_J(p)
 \Biggr] ,
\end{align}
where $p_z = (p_0-ip_1)/2$, $p_{\Bar{z}} = (p_0+ip_1)/2$,
and $*$ denotes the convolution
\begin{align}
 (\varphi_1 * \varphi_2)(p)
 \equiv \frac{1}{L_0L_1} \sum_q \varphi_1(q) \varphi_2(p-q).
\end{align}
The field products in~$\partial W(\{A\})/\partial A_I$
and~$\partial W(\{A\})/\partial A_I \partial A_J$ are understood as the convolution.

A remarkable property of the system is the existence of
the so-called Nicolai map~\cite{Nicolai:1979nr,Nicolai:1980jc}.
This mapping simplifies the path integral drastically;
the formulation makes essential use of it.
Now, we introduce new variables~$\{N\}$ as
\begin{align}
 N_I(p) &= 2i p_z A_I(p) + \frac{\partial W(\{A\})^*}{\partial A_I^*}(p) ,
 \label{eq:2.3}
\end{align}
which specify the Nicolai map from~$\{A\}$ to~$\{N\}$.
Note that the fermion determinant coincides with the Jacobian
associated with this mapping up to the sign.
After eliminating~$\{(\psi,\Bar\psi)\}$, the partition function is given by
\begin{align}
 \mathcal{Z}
 &= \int \prod_{|p_\mu|\leq\pi} \prod_I [d N_I(p) d N_I^*(p)]\,
 e^{- S_B}
 \sum_k \left.\sign \det
 \frac{\partial(\{N\}, \{N^*\})}{\partial(\{A\}, \{A^*\})} \right|_{\{A\}=\{A\}_{k}},
\end{align}
where $S_B$ is the bosonic part of the action,
$S_B = (1/L_0L_1) \sum_p \sum_I N_I^*(-p) N_I(p)$,
and $\{A\}_{k}$ ($k=1$, $2$, \dots) is a set of solutions of~eq.~\eqref{eq:2.3}.
The weight~$\exp(-S_B)$ is a Gaussian function of the variables~$\{N\}$.
To obtain configurations of~$\{N\}$ and~$\{A\}$,
we generate complex random numbers~$\{N(p)\}$ for each~$p_\mu$
from the Gaussian distribution,
and then, solve numerically the algebraic equation~\eqref{eq:2.3} with respect to~$\{A\}$.

The momentum-cutoff regularization, however, breaks the locality of the theory.\footnote{%
The present formulation is closely related
to the~$4$D lattice formulation~\cite{Bartels:1983wm}
based on the SLAC derivative~\cite{Drell:1976bq,Drell:1976mj}.
}
In the $2$D \emph{massive} WZ model,
one can argue the restoration of the locality in the continuum limit
within perturbation theory~\cite{Kadoh:2009sp}.
For the \emph{massless} case,
it is not clear whether the locality is automatically restored so far.
We believe that our numerical results support the validity of the present formulation.

\section{Numerical measurement of the central charge}
In a~$2$D SCFT, the central charge~$c$ appears in the two-point function of the EMT
\begin{align}
 \langle T(p) T(-p) \rangle
 &= L_0L_1 \frac{\pi c}{12} \frac{p_z^3}{p_{\Bar{z}}} ,
 \label{eq:3.1}
\end{align}
where the EMT, $T(p)=T_{zz}(p)$, is given in the momentum space by~\cite{Morikawa:2018ops}
\begin{align}
 T(p)
 &= \frac{\pi}{L_0L_1} \sum_q \sum_I \Bigl[
 4 (p-q)_z q_z A_I^*(p-q) A_I(q) \notag\\
 &\qquad\qquad\qquad\quad
 - i q_z \psi_{2I}(p-q) \Bar\psi_{\Dot{2}I}(q)
 + i (p-q)_z \psi_{2I}(p-q) \Bar\psi_{\Dot{2}I}(q)
 \Bigr] .
\end{align}
The IR behavior of the WZ model would be governed
by relations as~eq.~\eqref{eq:3.1} in SCFT.
The central charge can be computed
from the fit function~\eqref{eq:3.1} in the IR region.

\begin{figure}[t]
 \begin{minipage}{0.65\columnwidth}
  \begin{center}
   \begin{subfigure}{0.46\columnwidth}
    \begin{center}
     \includegraphics[width=\columnwidth]{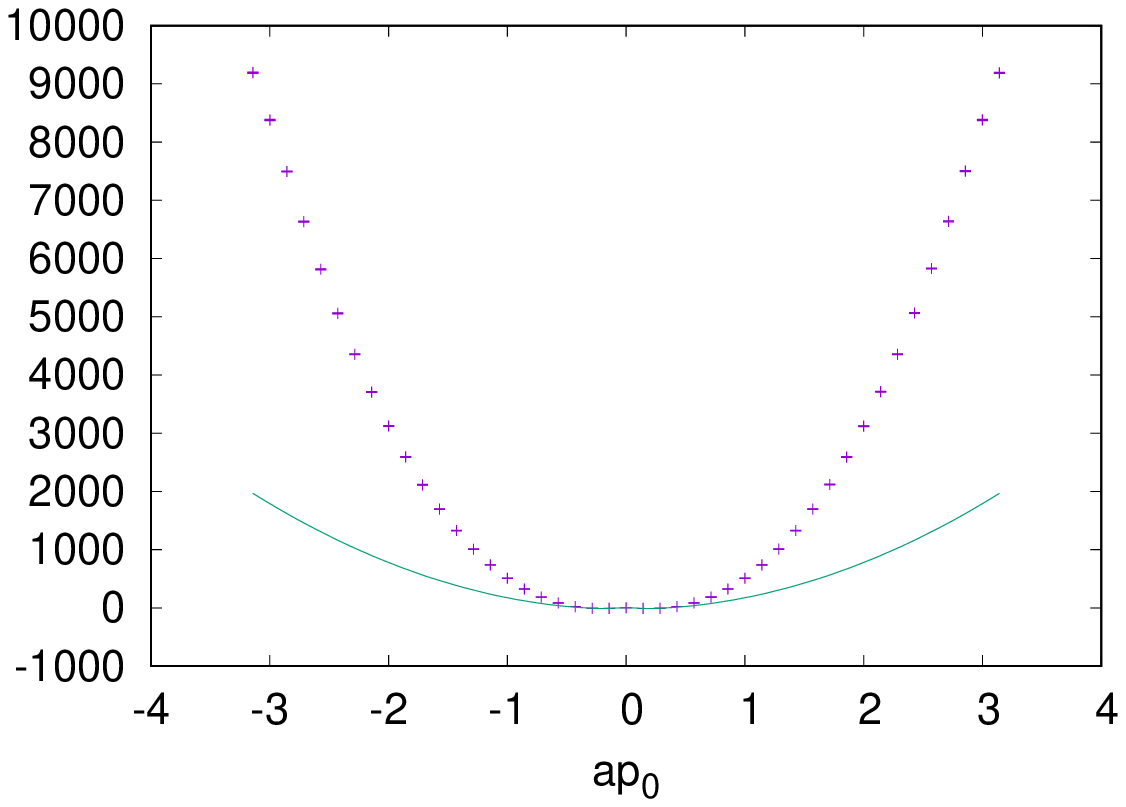}
     \caption{Real part}
    \end{center}
   \end{subfigure} \hspace*{1em}
   \begin{subfigure}{0.46\columnwidth}
    \begin{center}
     \includegraphics[width=\columnwidth]{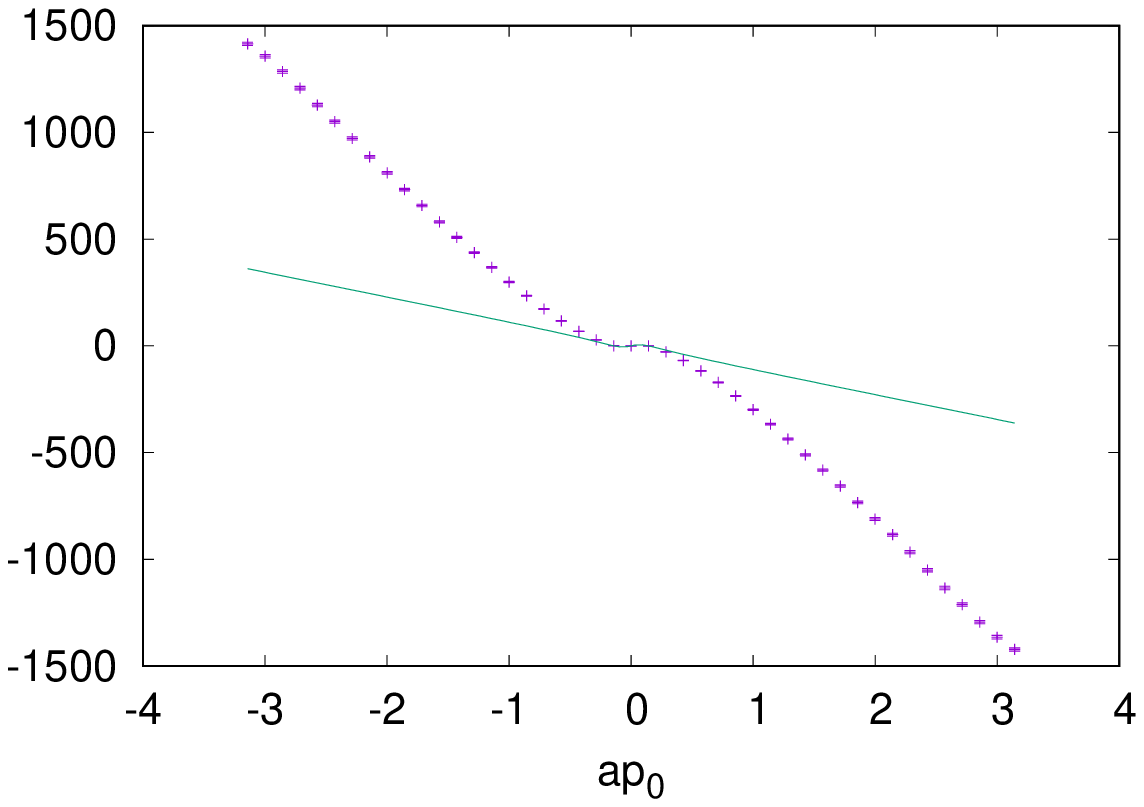}
     \caption{Imaginary part}
    \end{center}
   \end{subfigure}
  \caption{$\langle T(p)T(-p)\rangle$ for~$D_3$, $L/a=44$, and~$a p_1=\pi/22$.
  The fitting curve~\eqref{eq:3.1} is depicted at once.}
  \label{fig:emt_spt-d_k2L44}
  \end{center}
 \end{minipage}\hspace{1em}
 \begin{minipage}{0.3\columnwidth}
  \begin{center}
   \includegraphics[width=\columnwidth]{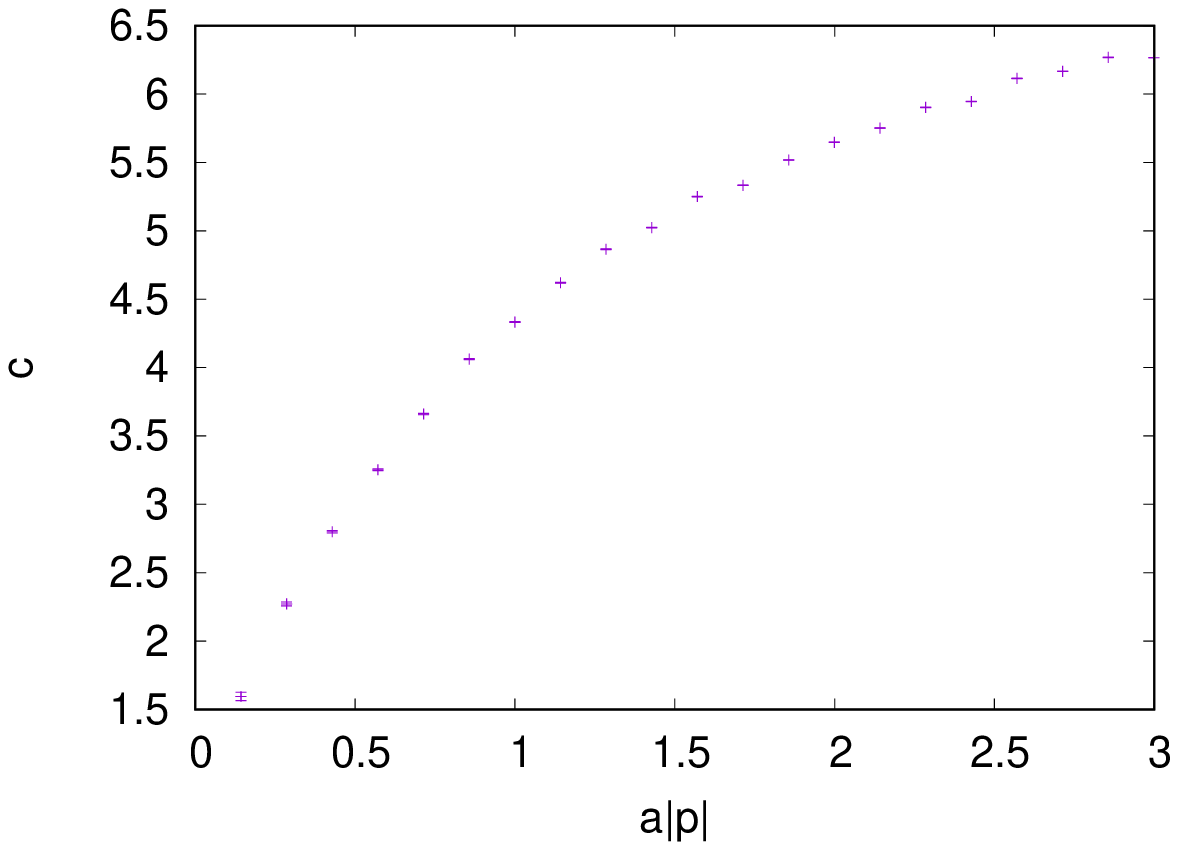}
  \caption{``Effective central charge'' for~$D_3$ and~$L/a=44$.
   }
  \label{fig:emt_fit}
  \end{center}
 \end{minipage}
\end{figure}

Let us show the first main result of this paper,
the numerical determination of the central charge
in the~$A_2$, $A_3$, $D_3$, $D_4$, and~$E_7$ models,
whose superpotentials are shown in~table~\ref{tab:ade};
for details of the computation, see~\cite{Morikawa:2018ops,Morikawa:2018zys}.
For the~$D_3$-type theory with~$L/a=44$, $a\lambda=0.3$ and~$a p_1=\pi/22$, for example,
we plot the correlation function~$\langle T(p)T(-p) \rangle$
in~figure~\ref{fig:emt_spt-d_k2L44} with the fitting curve~\eqref{eq:3.1};
the central charge~$c$ is obtained from the fit
in the IR region~$2\pi/L\leq|p|<4\pi/L$.
As is mentioned in~\cite{Kamata:2011fr,Morikawa:2018ops,Morikawa:2018zys},
it is interesting to plot the ``effective central charge,''
which changes as the function of~$|p|=2\pi n/L$ with fitted momentum regions,
$2\pi n/L\leq|p|<2\pi(n+1)/L$, for~$n\in\mathbb{Z}_{+}$;
it is analogous to the Zamolodchikov's~$c$-function.
Then figure~\ref{fig:emt_fit} shows that the ``effective central charge'' connects
the IR central charge to the UV one~$c=3N_{\Phi}$
in the expected free~$\mathcal{N}=2$ SCFT.

We tabulate the numerical results of the central charge
for the maximal box size for each setup in table~\ref{tab:c-res}.
These results are consistent with the expected values of the corresponding minimal models
within the numerical errors.
We have the numerical evidences of the following typical minimal models:
the $A_2$, $A_3$, $D_3$, $D_4$, $E_6$ ($\cong A_2\otimes A_3$), and $E_7$-type theories.

\begin{table}[t]
 \centering
 \begin{tabular}{lllll}\toprule
  Algebra & $L/a$ & $\chi^2/\text{d.o.f.}$ & $c$ & Expected value\\\midrule
  $A_2$ & 36 & 1.017 & 1.061(36)(34) & 1 \\
  $A_3$ & 30 & 0.916 & 1.415(36)(36) & 1.5 \\
  $D_3$ & 44 & 3.598 & 1.595(31)(41) & 1.5 \\
  $D_4$ & 42 & 1.177 & 2.172(48)(39) & 2 \\
  $E_7$ & 24 & 1.364 & 2.638(47)(59) & 2.666\dots \\
  \bottomrule
 \end{tabular}
 \caption{The central charge obtained from the fit of the EMT correlator
 with the maximal box size for each setup.
 The fitted momentum range is~$2\pi/L\leq|p|<4\pi/L$.
 Numbers in the second parentheses indicate the systematic error
 associated with the finite-volume effect given 
 in~\cite{Morikawa:2018ops,Morikawa:2018zys}.}
 \label{tab:c-res}
\end{table}%

\section{Continuum-limit analysis of the scaling dimension}
In the same way, we can also compute the scaling dimension~$h+\Bar{h}$
from the scalar correlator,
\begin{align}
 \left\langle A(x) A^{*}(0) \right\rangle = \frac{1}{z^{2h} \Bar{z}^{2\Bar{h}}},
 \label{eq:4.1}
\end{align}
for large~$|x|=\sqrt{x^2}$, where~$z=x_0+i x_1$, $\Bar{z}=x_0-i x_1$
and the conformal weights~$(h,\Bar{h})$ are supposed
to meet the spinless condition~$h=\Bar{h}$; see~table~\ref{tab:scaling_dim}.
It was found~\cite{Morikawa:2018ops} that,
although the measured scaling dimension tends to approach an expected value
as the grid size~$L/a$ increases,
the approach to the $L/a\to\infty$ limit appears not quit smooth.
To obtain a result in the continuum and the infinite volume,
we develop a systematic method of the continuum and thermodynamic limit.
To do this,
let us consider a numerical determination of the scaling dimension,
which is the finite-size scaling analysis given in~\cite{Kawai:2010yj}.
In this analysis, we observe the susceptibility of the scalar field~$A$, defined by
\begin{align}
 \chi(L_\mu)
 = \frac{1}{a^2} \int_{L_0 L_1} d^2 x\, \left\langle A(x) A^{*}(0) \right\rangle
 = \frac{1}{a^2L_0L_1} \left\langle |A(p=0)|^2 \right\rangle.
\end{align}
From the long-distance behavior~\eqref{eq:4.1},
we have the finite-volume scaling of the scalar susceptibility
for large~$L_\mu$, as~$\chi \propto (L_0 L_1)^{1 - h - \Bar{h}}$.
Numerically simulating the scalar correlator for some different volumes,
one can read the exponent, $1-h-\Bar{h}$,
from the slope of~$\ln\chi(L_\mu)$ as a linear function of~$\ln(L_0L_1)$.
In what follows, for simplicity, we set the physical box size~$L=L_0=L_1$,.

\begin{table}[t]
 \centering
 \begin{tabular}{lllll}\toprule
  Algebra & $L$ & $\chi^2/\text{d.o.f.}$ & $1-h-\Bar{h}$ & Expected value \\\midrule
  $A_2$ & 36 & 0.506 & 0.682(10)(7) & 0.666\dots \\
  $A_3$ & 30 & 0.358 & 0.747(11)(12) & 0.75 \\
  \bottomrule
 \end{tabular}
 \caption{Scaling dimension~$1-h-\Bar{h}$ obtained from the fit
 of the scalar correlator~\cite{Morikawa:2018ops}.}
 \label{tab:scaling_dim}
\end{table}

We develop this finite-volume scaling into an analysis method
with the continuum limit~\cite{Morikawa:2019cql}.
In what follows, for simplicity, we consider the~$A_n$-type LG model
with the superpotential, $W(\Phi)=\lambda\Phi^{n+1}/(n+1)$.
Our strategy of the continuum limit is as follows:
We regard~$\ln\chi(L)$ as the same kind of
the running coupling~$\Bar{g}^2(L)$ defined on a lattice~\cite{Luscher:1991wu}.
The lattice parameter~$a\lambda$ is tuned so that
$\ln\chi(L)$ is kept fixed; we set $\ln\chi(L)=u$.
Then, computing~$\ln\chi(2L)$ for~$2L/a$ and~$a\lambda$,
we observe the $a$-dependence of~$\ln\chi(2L)|_a$;
we denote $\Sigma(u, a/L) = \ln \chi(s L)|_a$,
where the statistical error of~$\Sigma$ is defined by a square root of the sum
of the squared errors of~$\ln\chi(L)$ and~$\ln\chi(2L)$.
With a to-be-determined fit function, the scaling dimension is given by
\begin{align}
 1 - h - \Bar{h}
 &= \frac{1}{\ln s^2} \left[ \lim_{a/L\to0} \Sigma(u, a/L) - u \right] .
 \label{eq:4.4}
\end{align}
To study the conformal behavior,
note that the unique mass scale~$\lambda$ in the~$A$-type theory
should be sufficiently larger than~$1/L$~\cite{Kawai:2010yj},
hence~$\lambda L\to\infty$ as the continuum limit~$a/L\to0$.
This implies that the above extrapolation method carries out the thermodynamic limit.
One can apply our continuum-extrapolation method to other lattice formulations,
e.g., that~in~\cite{Kawai:2010yj}.

Let us show the result of the precision measurement of the scaling dimension
for the~$A_2$-type theory with the cubic superpotential~$\Phi^3$;
for details of the computation, see~\cite{Morikawa:2019cql}.
From~table~$4$ in~\cite{Morikawa:2019cql},
we simply applies a linear function of~$a/L$ to~eq.~\eqref{eq:4.4}, then we have
\begin{align}
 1-h-\Bar{h} &= 0.6699(77)(87) ,
\end{align}
with~$\chi^2/\text{d.o.f.} = 1.417$.
This is the second main result in this paper.
Here, a number in the second parentheses indicates the systematic error
defined by the deviation between this central value
and a result with a slightly different fitted region; see~\cite{Morikawa:2019cql}.
This result is rather consistent with the expected exact
value~$1-h-\Bar{h}=2/3=0.6666\dots$ within the statistical error.

\section{Conclusion}
In this paper,
we numerically studied the IR behavior of the~$2$D~$\mathcal{N}=2$ WZ model
corresponding to the ADE minimal models,
by using the supersymmetry-preserving formulation
with the momentum cutoff~\cite{Kadoh:2009sp}.
First, we numerically measured the central charge of various typical minimal models:
$A_2$, $A_3$, $D_3$, $D_4$, $E_6$ ($\cong A_2\otimes A_3$),
and~$E_7$-type theories~\cite{Morikawa:2018ops,Morikawa:2018zys}.
Second, we gave the continuum-extrapolation method through the finite-size scaling
to determine the scaling dimension; then,
we performed the precision measurement of the scaling dimension~\cite{Morikawa:2019cql}.
Although the theoretical background of the formulation~\cite{Kadoh:2009sp}
is not clear so far,
our results are consistent with the conjectured correspondence
between the WZ model and the minimal series of SCFT,
and thus, support the validity of the approach.

For a possible application of the present numerical approach
to the Calabi--Yau compactification, the simulation of the LG theory
which corresponds to the~$A_4$ minimal model or a simpler non-minimal SCFT
will be an important starting point.

\acknowledgments
We are grateful to Daisuke Kadoh, Yoshio Kikukawa, Katsumasa Nakayama, Hiroshi Suzuki
and Hisao Suzuki for helpful discussions and comments.
The numerical computations were partially carried out by supercomputer system ITO
of Research Institute for Information Technology (RIIT) at Kyushu University.
This work was supported by JSPS KAKENHI Grant Number JP18J20935.

\end{document}